\documentclass[12pt]{iopart}

\usepackage{graphicx}
\usepackage{amssymb}
\usepackage{bm}
\usepackage[utf8]{inputenc} % Tell LaTeX that file is encoded in UTF-8
\usepackage[colorlinks=true]{hyperref}

\begin{document}
\title{Enhanced light emission from Carbon Nanotubes integrated in silicon micro-resonator}
\date{\today}

\author{Adrien Noury$^1$\footnote{Present address: ICFO-Institut de Ciencies
Fotoniques, Mediterranean Technology Park, 08860 Castelldefels, Barcelona,
Spain}, Xavier Le Roux$^1$, Laurent Vivien$^1$ and Nicolas Izard$^{1,2}$}
\address{$^1$ Institut d'Electronique Fondamentale, CNRS-UMR 8622, Univ. Paris-Sud, 91405 Orsay, France}
\address{$^2$ Laboratoire Charles Coulomb, CNRS-UMR 5221, Univ. Montpellier, 34095 Montpellier, France}
\eads{\mailto{adrien.noury@icfo.es}, \mailto{nicolas.izard@univ-montp2.fr}}

\begin{abstract}
Single-wall carbon nanotube are considered a fascinating nanomaterial for
photonic applications and are especially promising for efficient light emitter
in the telecommunication wavelength range. Furthermore, their hybrid
integration with silicon photonic structures makes them an ideal platform to
explore the carbon nanotube instrinsic properties. Here we report on the
strong photoluminescence enhancement from carbon nanotubes integrated in
silicon ring resonator circuit under two pumping configurations:
surface-illuminated pumping at 735 nm and collinear pumping at 1.26~$\mu$m.
Extremely efficient rejection of the non-resonant photoluminescence was
obtained. In the collinear approach, an emission efficiency enhancement by a
factor of 26 has been demonstrated in comparison with classical pumping
scheme. This demonstration pave the way for the development of integrated
light source in silicon based on carbon nanotubes.
\end{abstract}

\maketitle

\section{Introduction}
Carbon Nanotube photonics is an emerging field where researchers look for
their potential application in the framework of on-chip optical
communications.  Several elementary building blocks are required to achieve an
optical link.  Such a circuit should embed light emitters, optical modulators
and photodetectors coupled to an optical channel, in order to generate,
propagate and transmit informations encoded on light directly on a chip.
Thanks to their exceptional intrinsic optical
properties\cite{acs-Freitag,prb-Miyauchi,acs-Hertel}, single-walled carbon
nanotubes (SWNT) have the potential for building light emitting devices on
silicon.  Indeed, optical gain was demonstrated in polymer-wrapped
SWNT\cite{apl-Gaufres}, and Stark effect was observed in suspended
SWNT\cite{apl-Yoshida}.

In this field of nanotube photonics, a first milestone was to couple light
emission from SWNT into silicon waveguides \cite{acsnano-Gaufres,
adma-Khasminskaya}. Current research focuses on coupling SWNT with optical
cavities\cite{nature-Vahala, natnano-Xia, apl-Legrand, apl-Fujiwara,
ox-Gaufres} to enhance light-matter interaction. This scheme is based uppon
the work on the so-called Purcell effect\cite{pr-Purcell}, confirmed in
differents recent experiments of cavity quantum
electrodynamics\cite{Reithmaier2004, Sapienza2010}. Such an enhancement of
emission properties is highly desirable to overcome the low SWNT quantum
yield\cite{prb-Miyauchi}, and could be extended to reach strong-coupling
regime\cite{Peter2005} or to achieve laser emission\cite{Hill2004}.

Recently, several works focused on the integration of carbon nanotubes with
photonic cavities built on silicon-on-insulator (SOI) substrate, which is the
preferential platform for photonic applications. Such works included coupling
nanotubes with photonic crystal suspended cavity\cite{apl-Watahiki,
apl-Sumikura} or microdisk resonators\cite{apl-Imamura}, and high coupling
efficiency with a nanobeam cavity was recently reported\cite{natcom-Miura}.
However, these kinds of cavities were addressed using an out-of-the-plane
micro-photoluminescence configuration, which hinders efficient integration
into an optical link.

Building up on our previous work\cite{nanotech-Noury, KatoView2015}, we propose to couple
carbon nanotube photoluminescence (PL) with silicon microring resonators, in a
fully integrated configuration with an access waveguide. Using this
integration scheme, we show that it is possible to collect PL coupled to ring
modes through the access waveguide, with efficient rejection of non-resonant
photons. Emission quality factors almost up to 8000 were observed: they are
the highest values reported so far for carbon nanotubes coupled with an
optical cavity. Moreover, this design enables to collinearly excite SWNT
through the same access waveguide, leading to efficient excitation and
collection of SWNT PL at different wavelengths. We obtained an emission
enhancement of $26$ in this configuration, compared to the externally-pumped
configuration. The requirement of an external laser source for
out-of-the-plane excitation of SWNT is lifted, allowing pumping from an
integrated laser source and above all underlining the pertinence of this
flexible approach for realistic carbon nanotube based photonic devices.

\section{Methods}
\subsection{Ring resonator fabrication}
\begin{figure}
	\centering
	\includegraphics[width=12cm]{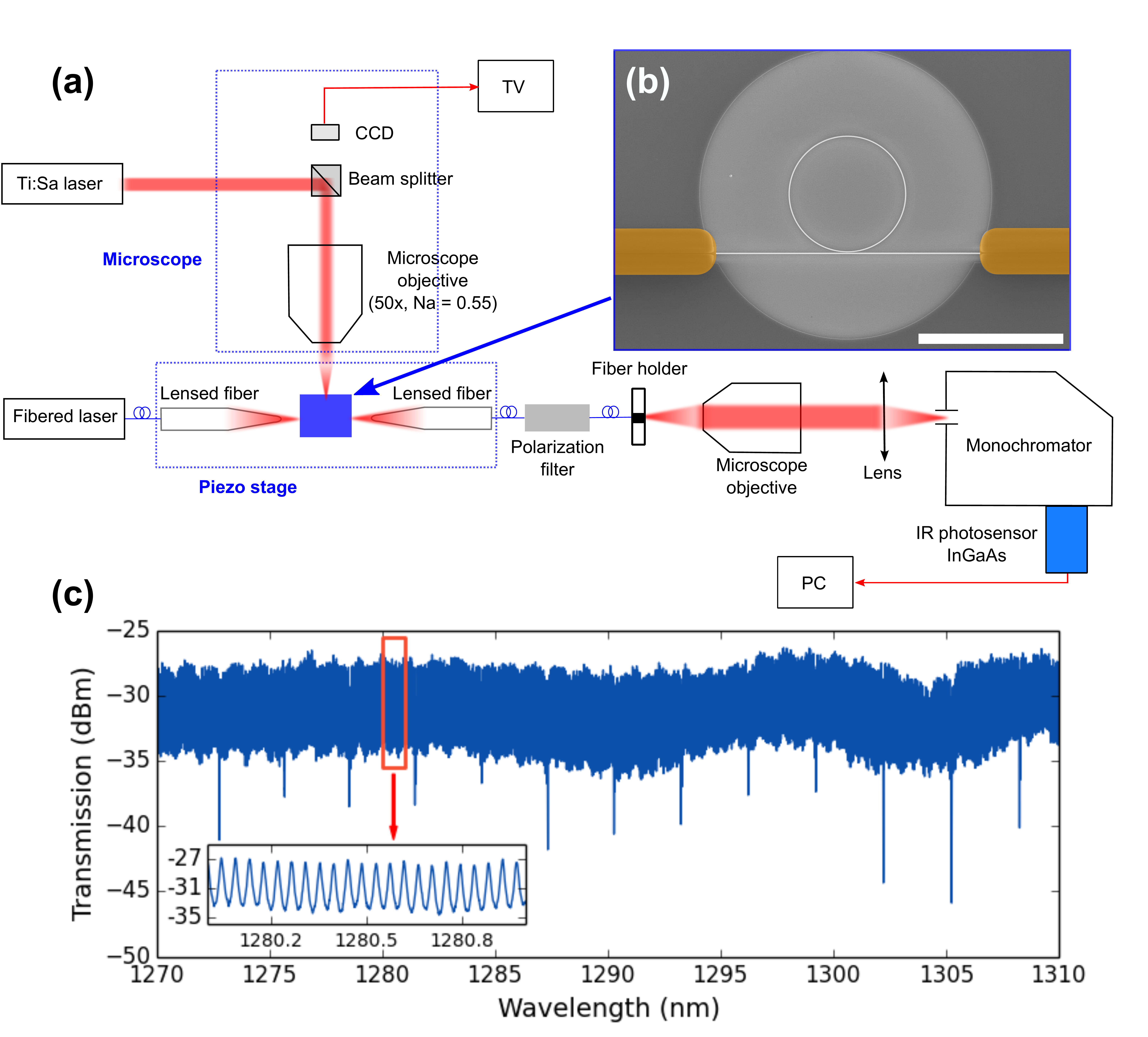}
	\caption{
(a) Characterization set-up for the ring resonators covered with SWNT. The
setup allows measuring the rings in two different configurations: externally
excited photoluminescence and collinearly excited photoluminescence. Only the
input changes in these configurations, while the collection system stays the
same. The SEM picture (b) of the ring shows the HSQ cover and both the bus and
the ring waveguides. Scale bar is 50~$\mu$m. The picture was taken prior to
SWNT deposition. False colors indicate the silicon ring and waveguide (white),
HSQ cladding (orange), silica (light grey) and silicon slab (dark grey). (c)
Typical transmission spectra of a microring resonator prior to SWNT
deposition. Here the ring radius was 20~$\mu$m and the coupling distance was
approximately 120~nm. The width of the waveguide was 340~nm and is considered
single-mode at the wavelengths of this study.
}
	\label{fig1}
\end{figure}

Photonic structures are made from SOI substrates, with a 220~nm silicon layer
and a 2~$\mu$m thick buried silicon oxyde. Waveguides and photonic cavities
are defined by electron beam lithography followed by dry etching using an
inductively coupled plasma. The typical width of microring resonators is
340~nm. An additional lithography step was performed to form a local cladding
(height $\approx 900~nm$) for waveguides, using Hydrogen SilsesQuioxane (HSQ).
After patterning, HSQ was processed into a silica-like layer, in order to
achieve mode symmetrisation and waveguides isolation from carbon nanotubes
outside well-defined interaction regions. A typical SEM picture of a microring
resonator (prior to nanotube deposition) coupled to an access bus waveguide is
displayed in figure~\ref{fig1}(b).

\subsection{SWNT extraction and deposition}
Single-wall carbon nanotube are synthesized using the HiPCO process and
obtained from a commercial source (Unydim). Semiconducting SWNT were extracted
using a polyfluorene (PFO) based process\cite{Nish2007, Chen2007, apl-Izard},
known to yield to high-purity material, with no visible trace of metallic
nanotubes and a handful of remaining semiconductor SWNT
chiralities\cite{apl-Izard,ol-Gaufres}.  Among them are the (8,7) and (9,7)
SWNT species, with PL emission located within the telecom O Band, respectively
centered at 1280~nm and 1350~nm. Therefore, this material is perfectly
suitable for integration in an optical interconnect scheme based on silicon,
at telecommunication wavelengths. The extracted material was then spin-casted
on top of the sample at 1000~RPM during 60~s. This procedure was repeated
three times to ensure the deposition of a thin and flat layer. To further
increase the quantity of carbon nanotubes on the sample and in order to
achieve better symmetrization of the optical mode in the exposed interaction
region, a drop-casting was carried out. Finally, a thermal annealing at 180
$^{o}$C during 15~min was used to cure remaining inhomogeneities in the
deposited layer.

\subsection{Integrated PL setup}
The characterization setup is represented in figure~\ref{fig1}(a) and is
composed of a tunable fiber laser source used for inputing light into the
waveguides, and a Ti:Sapphire external laser source used to pump the SWNT on
their second excitonic transition S$_{22}$, focused on the sample by a
microscope objective (50x, NA~=~0.55). The collection system is made using a
polarization maintaining fiber, a polarization filter and a spectrometer
composed of a 950~lines/mm~grating coupled to a nitrogen-cooled InGaAs
photodetector array. Depending on the experiment, the optical pump could be
either the fiber laser or the Ti:Sapphire source.

\section{Results and Discussion}
\subsection{Transmission measurements}
The ring resonator transmission without carbon nanotubes on the surface was
recorded. To ensure that the symmetrization of the optical mode is similar
with or without presence of nanotubes, we used a liquid with a well-known
refractive index ($n = 1.46$), close to the expected index for the
PFO-embedded carbon nanotubes material ($n \approx 1.5$\cite{Campoy2005}). The
experimental aparatus is similar to the one depicted schematically in
figure~\ref{fig1}(a). The set-up to measure transmission consists of a tunable
fiber laser source, with a polarization control system and a lensed fiber for
injection of the light into the chip. The output light is collected by a
microscope objective, filtered in polarization and monitored by a
photodetector.

An exemple of transmission spectrum is displayed in figure~\ref{fig1}(c).
Injection power was 1~mW for TE polarization. The studied structure was a
microring resonator with a 20~$\mu$m radius. The input light from the access
waveguide evanescently couples to the microring resonator and, depending of
the wavelength, is able to excite a resonance mode of the ring, leading to
this periodically spaced transmission dips. Microring resonators are
caracterized by the interval between two adjacent dips, namely the Free
Spectral Range (FSR), which is given by :
$$\Delta \lambda = \frac{\lambda_{1} \cdot \lambda_{2}}{n_{g} \cdot L}$$
where $\lambda_{1/2}$ are the center wavelength of the two adjacent dips
considered, $n_{g}$ the group index of the mode ($n_{g} \approx 4.2$ at 1300
nm for TE polarization) and the cavity length is $L = 2 \pi R$, with $R$ the
ring radius. On Fig~\ref{fig1}c, the FSR was estimated to $3.06 \pm 0.13$~nm
around 1.3~$\mu$m for a microring of 20~$\mu$m radius, which is consistent
with the expected group index of $n_{g} \approx 4.2$. Note that the FSR error
bar given represents the fit error. The oscillations observed between the
cavity modes were Fabry-Pérot fringes, as visible in the inset in
figure~\ref{fig1}(c). They originate in the reflection at the sample facets
and are not noise in nature.

\subsection{Externally pumped PL}
By using a Ti:Sapphire external laser source, SWNT could be resonantly excited
on their $S_{22}$ excitonic transition. The setup is represented in
figure~\ref{fig1}(a), where the Ti:Sapphire laser source is focused on the
sample using a 50x microscope objective. Light emitted by carbon nanotubes
(over their $S_{11}$ excitonic transition) excites a resonance mode of the
microring and evanescently couples to the access waveguide. Light was then
collected at the waveguide output by a lensed optical fiber. Additionnaly, a
polarizer enables to discriminate either TE or TM polarizations at the output.

\begin{figure}
	\centering
	\includegraphics[width=12cm]{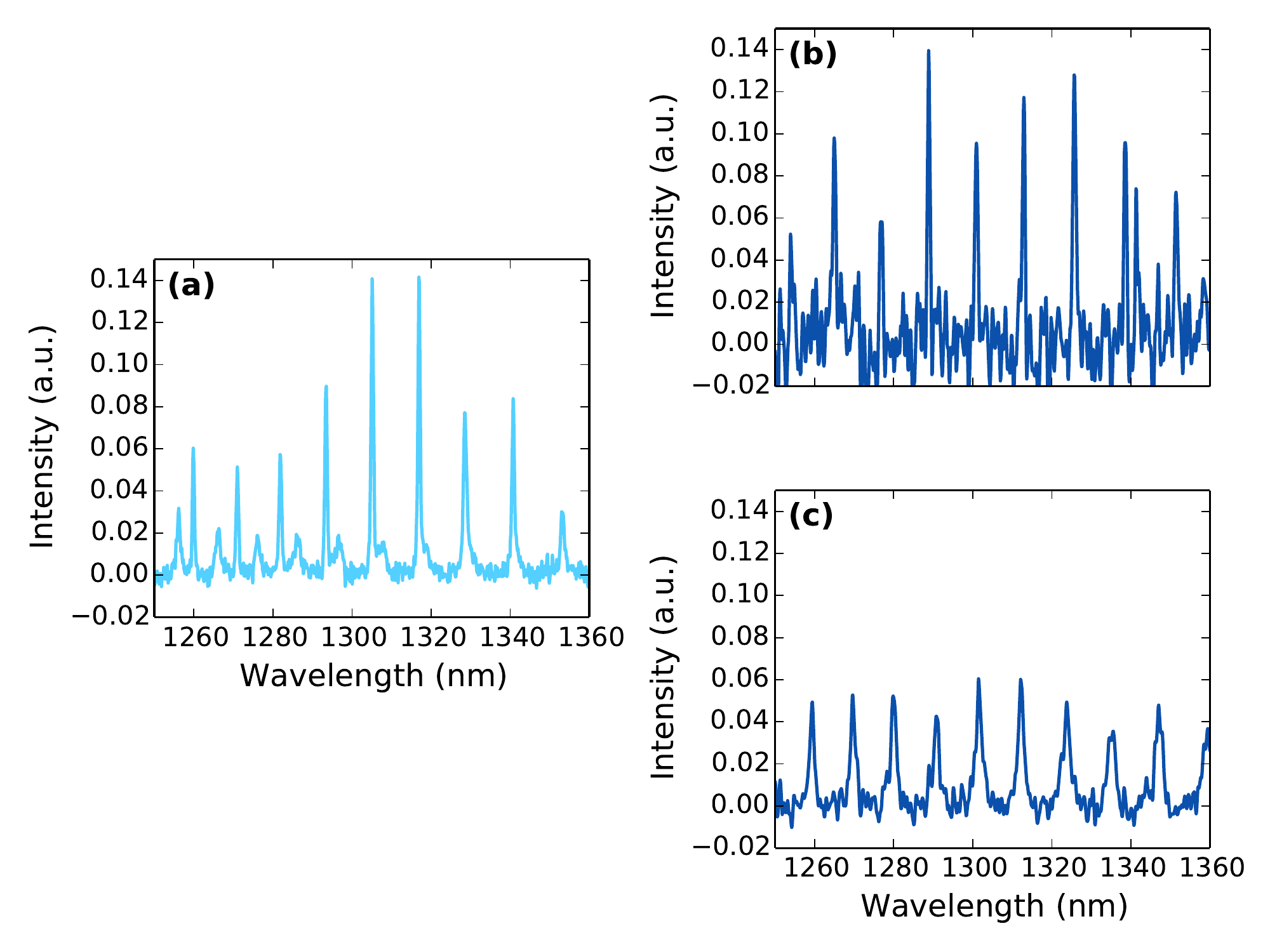}
	\caption{
Photoluminescence spectra collected at the waveguide output, using Ti:Sapphire
excitation, on a ring with radius $R = 5~\mu m$. (a) unpolarized output, (b)
TE polarized output and (c) TM polarized output. PL intensities are normalized
and represented on a same scale, for comparison. In each case, we used a
Lorentzian fit of the various resonance peaks to extract FWHM and FSR.
}
	\label{fig2}
\end{figure}

The figure~\ref{fig2} displays the three possible collected PL spectra : (a)
without polarization filter, (b) TE-filtered and (c) TM-filtered.

On each spectrum, evenly spaced narrow peaks are visible, corresponding to
ring resonance modes excited by nanotube photoluminescence. Interestingly, the
SWNT broadband PL background\cite{nanotech-Noury} is not observed in this
configuration, which we attribute to the intensity ratio of microring resonant
mode. Indeed, on the transmission spectra displayed in figure~\ref{fig1}(b),
the extinction ratio is as high as 13~dB. Thus, this configuration allows to
efficiently reject the uncoupled SWNT PL background, keeping only resonant
photons with high spectral purity.

\begin{table}
\begin{center}
\begin{tabular}{|c|c|c|}
	\hline
	& \textbf{FWHM} (nm) & \textbf{FSR} (nm)\\
	\hline
	Unpolarized output (set 1) & 0.67 $\pm$ 0.09 & 11.56 $\pm$ 0.48\\
	Unpolarized output (set 2) & 1.26 $\pm$ 0.31 & 10.54 $\pm$ 0.70\\
	\hline
	TE polarized output & 0.71 $\pm$ 0.14 & 12.08 $\pm$ 0.62\\
	TM polarized output & 1.41 $\pm$ 0.22 & 10.83 $\pm$ 0.56\\
	\hline
\end{tabular}
\end{center}
\caption{
Full-Width Half-Maximum (FWHM) and Free Spectral Range (FSR) from resonance
peaks of PL spectra in Fig.~\ref{fig2} (ring radius $R = 5~\mu m$), determined
using Lorentzian deconvolution. Indicated error is the fit standard deviation.
}
\label{tab1}
\end{table}

Unpolarized spectrum (figure~\ref{fig2}(a)) clearly shows two peak sets that
are not observed on both polarized spectra (figures~\ref{fig2}(b) and (c)).
Unpolarized and polarized spectra were deconvoluted with a Lorentzian fit to
extract the Full-Width Half-Maximum (FWHM) of the peaks and the FSR.
Corresponding data are summarized in table~\ref{tab1}.

The extracted FWHM on the polarized output is 0.71$\pm$0.14~nm for the TE
polarization and 1.41$\pm$0.22~nm for TM polarization. On the other hand, the
unpolarized PL spectrum clearly shows that the two peak sets have differents
FWHM. The deconvoluted narrow set had a FWHM of 0.67$\pm$0.09~nm and was
attributed to the TE polarization modes, while the broader set had a FWHM of
1.26$\pm$0.31~nm and was attributed to the TM polarization. Thus, we conclude
that the two peak sets observed in the unpolarized spectrum arise from both
fundamental polarizations, TE and TM, propagating in the microring resonator,
as previously suggested\cite{nanotech-Noury}.

We also notice that both TE and TM peak sets have a slightly different FSR.
Estimated FSR range from 10.5 to 12 nm, with a FSR slightly larger for TE
polarization, and are consistent between unpolarized and polarized output.
The observed correspondance between FSR in each type of polarization
experiment supports our attribution of the TE and TM modes in the unpolarized
PL spectrum.

\begin{figure}
	\centering
	\includegraphics[width=12cm]{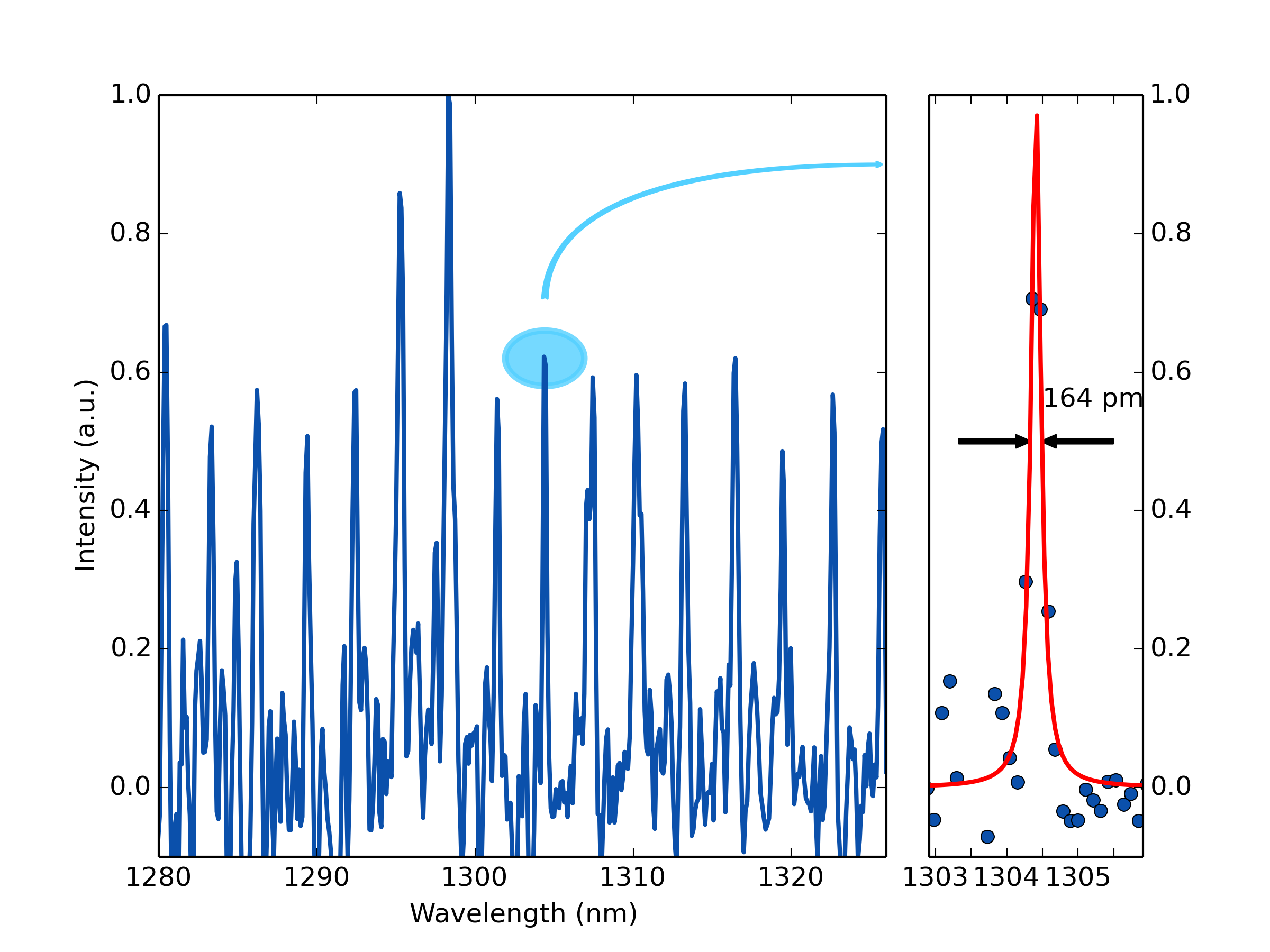}
	\caption{
(left) SWNT photoluminescence spectrum on a 20~$\mu$m radius microring
resonator and (right) magnified spectrum of the emission peak centered at
1304.4 nm. Lorentzian adjustment of the displayed peak gives a FWHM of 164 pm,
corresponding to a nearly 8000 quality factor.
}
	\label{fig3}
\end{figure}

An interesting feature of ring resonators is that their quality factor scales
with the radius\cite{Peccianti2012}, due to the reduction of the bending loss
with higher ring radius. PL spectra were recorded from nanotubes coupled into
microring resonators with radius up to 20~$\mu$m. The figure~\ref{fig3} shows
an exemple of PL spectra for a ring resonator with a radius of 20~$\mu$m. Very
sharp peaks are observed in this structure. In this case, the measured FWHM
was 164 pm, corresponding to a quality factor $Q$ of approximately 8000. This
value is the highest reported up to date, doubling the reported quality
factor\cite{apl-Imamura,nanotech-Noury,natcom-Miura} so far. Moreover, the
measured quality factor is suspected to be limited by the resolution of our
acquisition setup, and could probably be even higher.

\subsection{Spatial cartography of excitation}
\begin{figure}
	\centering
	\includegraphics[width=12cm]{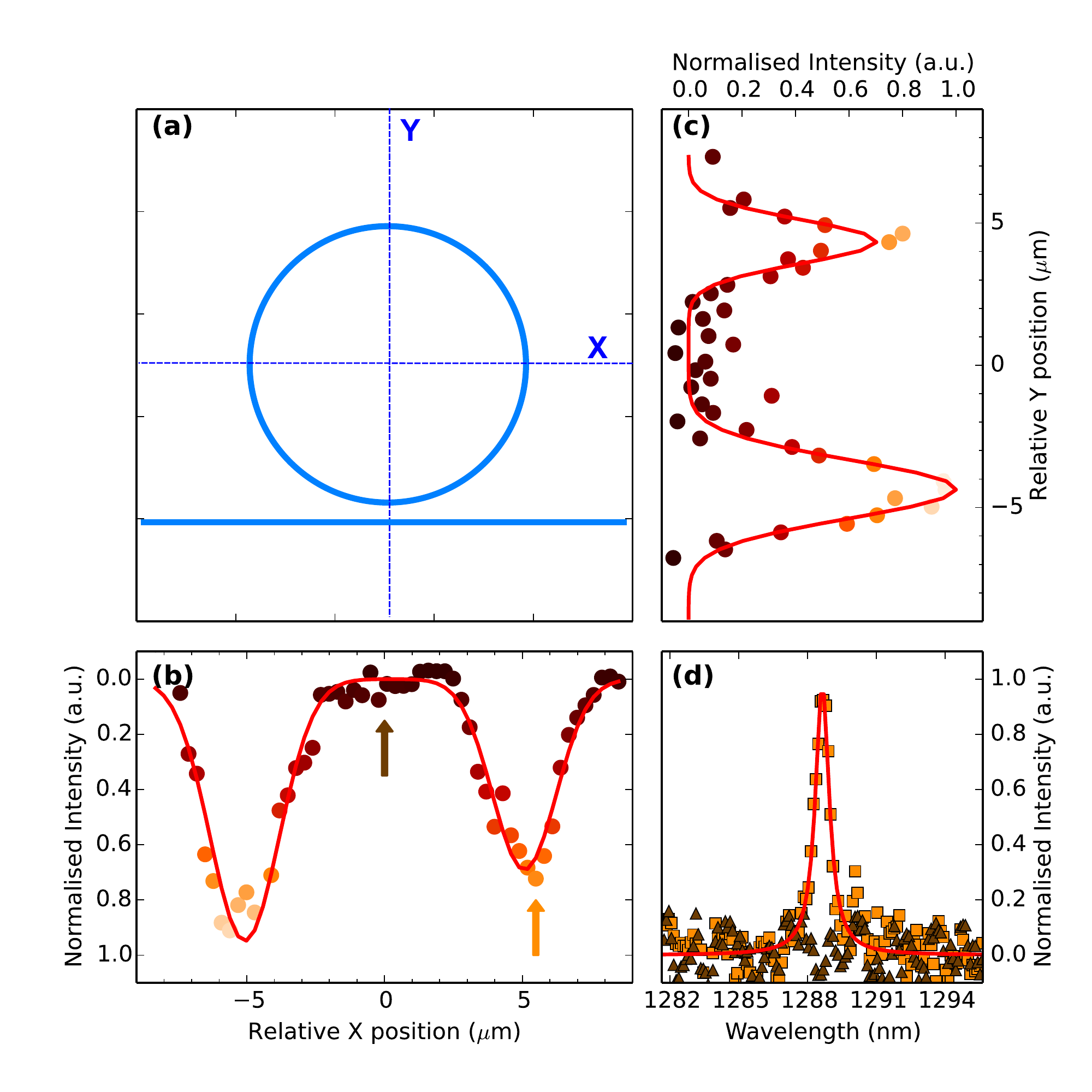}
	\caption{
Spatial cartography of the photoluminescence observed at the waveguide output.
(a) schematic of the displacement of the excitation beam over the structure.
(b) integrated intensity of a 1~nm wide window centered at 1288.66~nm, while
displacing the beam parallel to the bus waveguide on the X coordinate.  (c)
same as in (b) with displacement perpendicular to the bus waveguide on the Y
coordinate. (d) typical PL spectra on microring resonator, with the 1~nm wide
integration window, on the right side of the microring (square, orange), and
far from the ring wall (triangle, brown). Spectra displayed in (d) were
recorded at positions respectively indicated by the orange and brown arrows.
	}
	\label{fig4}
\end{figure}

The integrated measurement set-up (sample and fibers) was assembled on a piezo
stage controller in order to record a spatial cartography of the
photoluminescence. Such an experiment was performed by moving the excitation
beam over the ring resonator while recording a photoluminescence spectra in a
narrow window, in order to minimize the drift in the collection fiber
position. Two PL spectra, corresponding to a beam located on the right side of
the ring (square, orange) and 5~$\mu$m away from the ring (triangle, brown),
are displayed in figure~\ref{fig4}(d). A 1~nm wide spectral window was
integrated over the displayed peak. The results, as a function of spatial
coordinates, are reported on both axes in figures~\ref{fig4}(b) and (c).

Two maxima are observed on each plot and are attributed to the overlap between
the gaussian profile of the excitation laser and the gaussian profile of the
studied resonance mode of the microring resonator. It is reasonable to
consider that the 340~nm wide waveguide has a negligible size compared to the
pump laser beam waist, i.e. that the mode extension\cite{acsnano-Gaufres} of
the microring is much smaller than the laser beam extension, $W_{ring} <<
W_{beam}$. Therefore, it is possible to recover the beam profile by directly
fitting the two observed peaks. The beam profile was determined to be
$W_{beam}$=$2.8 \pm 0.2~\mu$m. The spacing between these two peaks fits with
the microring diameter, $D$=$10~\mu$m, emphasizing the fact that the observed
sharp peaks on the PL spectra originate from the microring cavity modes.
Moreover, this spatial cartography indicates that SWNT excitation is
homogeneous on different quadrants. This means that the microring mode
excitation did not rely on some hot-spot, but rather that the excitation is
delocalized on the whole microring length.

Interestingly, the spectra for a pump beam several microns away from the ring,
displayed in figure~\ref{fig4}(d), shows no trace of the SWNT PL.
Nevertheless, it is reasonable to consider that the spot is located on some
carbon nanotubes, since our deposition method does not localize nanotubes only
on the ring walls. We can then conclued that the nanotubes excited here do not
couple to the ring resonator. Therefore, this observation reveals that when
such coupling occurs, it is only mediated by near-field interaction between
SWNT - acting as a light emitting dipole - and the optical modes of the ring
resonator.

\subsection{Degenerated PL via collinear excitation}
To demonstrate the full potential of SWNT in an integrated configuration, we
took advantage of the microring configuration to pump SWNT through the access
waveguide. The Ti:Sapphire laser source was replaced by a fibre laser source.
In this configuration, the optical pumping was achieved above the first SWNT
excitonic transition S$_{11}$, since the silicon access waveguide and
microring could only transmit near-IR light.

\begin{figure}
	\centering
	\includegraphics[width=12cm]{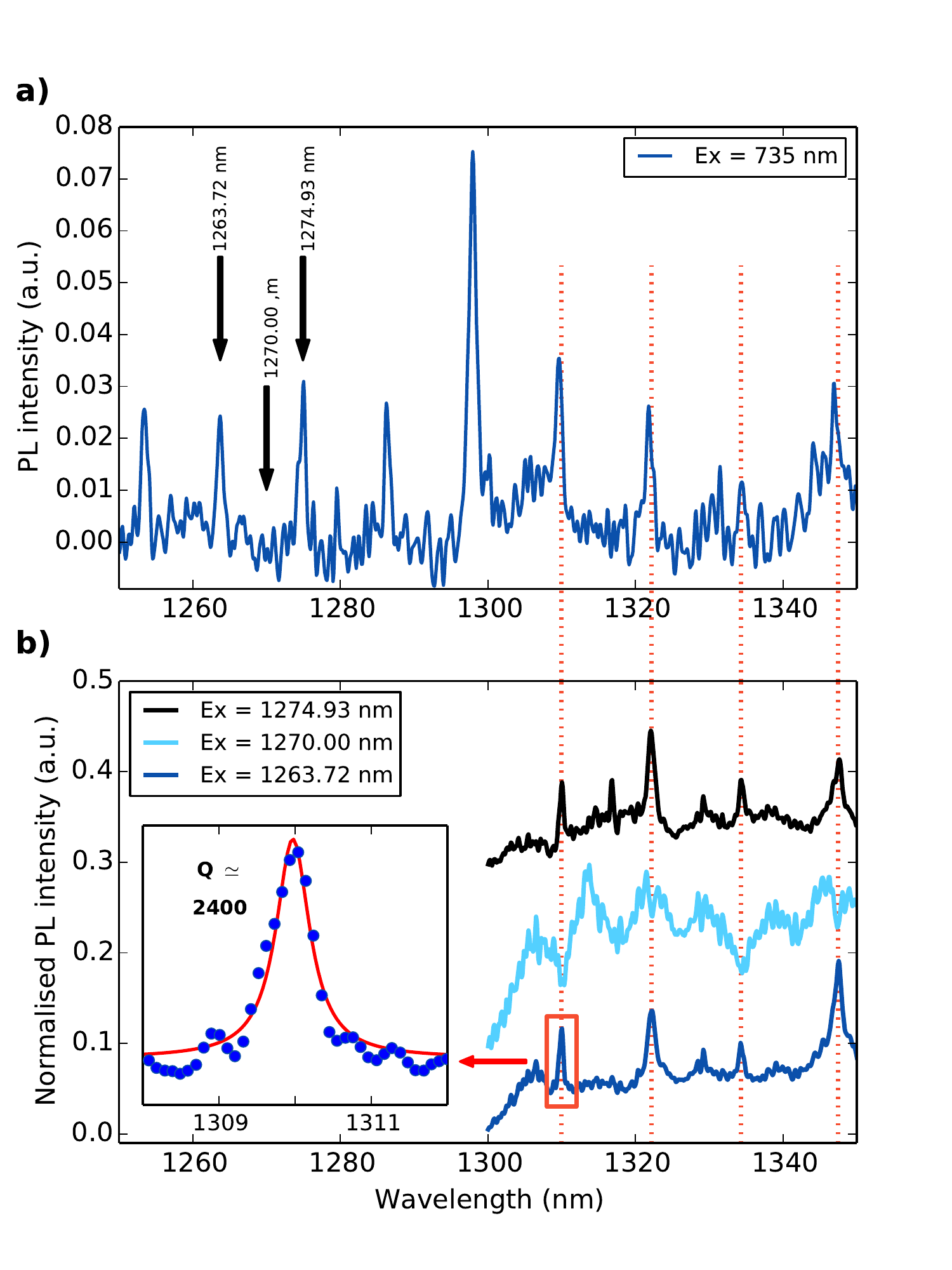}
	\caption{
(a) Reference PL spectra (TE polarized detection), recorded with Ti:Sapphire
excitation on SWNT second excitonic transition $S_{22}$ at 735~nm. This
spectrum is used to identify the resonance wavelengths of the microring
resonator (excited by SWNT PL) and to choose three pumping wavelengths, marked
by arrows, for collinear pumping experiments using the fiber laser source. (b)
PL spectra recorded with collinear TE polarized excitation by the fiber laser
light propagating through the access bus waveguide. Spectra were measured with
different pump wavelengths, then were normalized to unity and upshifted for
clarity. Inset displays a typical resonance in this configuration, with a
Lorentzian adjustment as a solid line.
  }
	\label{fig5}
\end{figure}

First, the pumping wavelength was selected by looking at a Ti:Sapphire excited
PL spectrum, displayed in figure~\ref{fig5}(a), recorded in the same
conditions as previously shown (TE-polarized output). Wavelengths of interest
for integrated pumping (TE polarized pump laser) were identified: two of them
were chosen in resonance with the microring mode, respectively at 1263.72~nm
and 1274.93~nm, while one was chosen out of resonance at 1270.00~nm. When
setting such wavelengths with the pump fibre laser, close to the acquisition
window, it is necessary to reject the transmitted pump residual. This is
achieved by placing a free-space low-pass filter with a cut-off wavelength of
1300~nm in front of the entrance of the spectrometer.

PL spectra recorded at the waveguide output using these three different
excitation wavelengths are displayed in figure~\ref{fig5}(b), for TE
polarization.

When the pumping wavelength was matching a resonance of the microring, either
at 1263.72~nm or 1274.93~nm (dark blue and black plots in
figure~\ref{fig5}(b)), the PL spectra displayed evenly spaced peaks. The
locations of these photoluminescence peaks are in agreement with the reference
spectra, as highligted by the dotted lines. We then conclude that these peaks
originate from the ring resonance modes excited by the nanotubes.

The fact that these peaks were absent in the case of the non-resonant pump
(light blue in figure~\ref{fig5}(b)) allows to understand precisely the
mecanism involved here: light inputed through the waveguide evanescently
couples in the latter. This coupled light excited nanotubes just above their
first excitonic transition. If the pump wavelength matches a resonance of the
ring resonator, the excitation is much more efficient due to the long photon
lifetime time in the resonator. After relaxing to the first excitonic state,
part of the energy was emitted back by the nanotube through photons, the
number of emitted photons being controlled by the nanotubes internal quantum
yield. The emitted photons coupled to lower energy resonance modes of the
microring (i.e., higher wavelength). These excited resonance peaks of the ring
then evanescently coupled to the bus waveguide and were collected at the
sample output.

Noticeably, some dips are also observed in figure~\ref{fig5}(b) in the case of
non-resonant pump, matching the resonance peaks highlighted by the dotted
lines. This is attributed to the remnant of the pump laser, not completely
extinguished by the low-pass filter and responsible for the broad background
observed, which is filtered by the microring cavity mode.

We finally note that, in this configuration, the output light intensity was
higher compared to the situation where the PL spectra are excited by the
Ti:Sapphire. Indeed, if we compare the emission area of the peak centered at
1347 nm and normalize it by the respective excitation pump power, we find an
area of 529.3 photons/W and 20.1 photons/W for collinear pumping
($\lambda_{Pump} = 1274.93~nm$) and for Ti:Sa pumping ($\lambda_{Pump} =
735~nm$) respectively. This leads to an increase of the emission by a factor
higher than 26, which is attributed to the more efficient excitation scheme
involved when pumping nanotubes by the resonance mode of the ring, rather than
exciting them by an external pump. Again, this shows the strength of such a
fully-integrated approach to excite SWNT PL.

\section{Conclusion}
In conclusion, we have demonstrated the coupling of carbon nanotubes
photoluminescence with silicon microring resonators in a fully integrated
configuration. This powerful approach make possible the collection of SWNT PL
at a long distance after the microring resonator, enabling information
transmission within an integrated optical link. Plus, the integrated approach
reported here led to the elucidation of the polarization effect occuring in
microring resonators that were previously suspected\cite{nanotech-Noury}. It
was demonstrated that the fundamental polarizations TE and TM could propagate
in the microring resonator, with a slightly different FSR.

A spatial cartography of the photoluminescence was also performed,
demonstrating that SWNT excitation is delocalised on the whole microring
length and did not originate from an hot-spot, but was based on near-field
coupling between SWNT and the optical modes of the microring.

The quality factor $Q$ of the emission peaks achieved here reached up to 8000
for 20~$\mu$m radius microring, doubling the previously reported values.
Finally, we demonstrated the successful pumping of SWNT through the access
waveguide, displaying a higher overall emission, compared to the externally
pumped configuration, with an increase of emission by a factor more than 26.

Pumping on the first excitonic transition of SWNT is easier to implement
because it lifts the requirement of an external source and takes advantage of
the silicon transparency to propagate the excitation wave. We demonstrated
that such an excitation path is highly efficient in microring resonators if
realized on a higher energy resonance of the ring, as reported in other types
of cavities\cite{apl-Imamura, Liu2015}.

As quality factor of ring resonator scales with the ring radius, it is
expected that the measured quality factor can be greatly improved by
increasing the ring dimensions. Combining high-Q cavity with simple design, an
efficient excitation scheme, and a material displaying optical
gain\cite{apl-Gaufres} is expected to lead to the demonstration of SWNT based
laser.

The analysis of the different pumping schemes, on/off resonance, allowed us to
understand the mechanisms of the interaction between the optical mode of the
resonator and SWNT deposited on the microring. Such a precise understanding is
valuable for the design of more efficient interaction schemes for SWNT or any
material in photonic cavity.

\section*{Acknowledgments}
A. Noury acknowledges the Ministry of Higher Education and Research (France)
for scholarship. The authors acknowledge A. Degiron for assistance and
fruitful discussions. This work has been supported by ANR JCJC project
"\c{C}a~(Re-)~Lase~!", by the Region Ile-de-France in the framework of DIM
Nano-K (project CANOA) and by the FET project "Cartoon". Fabrication was
performed in the IEF clean room facilities (CTU/MINERVE), part of the RENATECH
network.

\section*{References}

\bibliography{microring}{}
\bibliographystyle{unsrt}

\end{document}